\documentstyle[12pt]{article} 
\newcommand{\beq}{\begin{equation}} 
\newcommand{\eeq}{\end{equation}} 
\newcommand{\bea}{\begin{eqnarray}} 
\newcommand{\eea}{\end{eqnarray}} 
 
\newcommand\refpar[1]{(\ref{#1})}

\newcommand{\J}{\hbox{${\bf J}$}}
\newcommand{\x}{\hbox{${\bf x}$}}
\newcommand{\pv}{\hbox{${\bf v}$}}
\newcommand{\Vs}{\hbox{${\bf V}_s$}}
\newcommand{\vs}{\hbox{${\bf v}_s$}}
\newcommand{\vi}{\hbox{${\bf v}_{{\rm i}}$}}
\newcommand{\Vn}{\hbox{${\bf V}_n$}}
\newcommand{\vn}{\hbox{${\bf v}_n$}}
\newcommand{\Us}{\hbox{${\bf U}_s$}}
\newcommand{\wus}{\hbox{$\widetilde{{\bf U}}_s$}}
\newcommand{\kzerodeux}{\hbox{${\bf k}_{02}$}}
\newcommand{\kzeroun}{\hbox{${\bf k}_{01}$}}
\newcommand{\f}{\hbox{${\bf f}$}}
\newcommand{\bomega}{\hbox{{\boldmath $\omega_s$}}}
\newcommand{\wbomega}{\hbox{$\widetilde{{\boldmath \omega}}_s$}}
\newcommand{\wbomegap}{\hbox{$\widetilde{{\boldmath \omega}}_{s \perp}$}}
\newcommand{\W}{\hbox{${\bf W}$}}
\newcommand{\q}{\hbox{${\bf q}$}}
\newcommand{\diff}{{\rm d}}
\newcommand{\tc}{\hbox{$T_{\lambda}$}}
\newcommand{\vecinc}{\hbox{$\hat \imath$}}

\title{Scattering of first and second sound waves by quantum vorticity  
in superfluid Helium} 
 
\author{{\sc Christophe Coste} \\ Laboratoire de Physique, \'Ecole Normale  
Sup\'erieure de Lyon \\ 46, All\'ee d'Italie,69364 Lyon Cedex 07, France 
\and 
{\sc Fernando Lund} \\Departamento de F\'\i sica, Facultad de Ciencias 
 F\'\i sicas y Matem\'aticas \\ Universidad de Chile, Casilla 487-3,  
 Santiago, Chile} 
 
\date{} 
 
\begin{document} 
 
\maketitle 
 
\begin{abstract} 
We study the scattering of first and second sound waves by quantum 
vorticity  
in superfluid Helium using two-fluid hydrodynamics. The vorticity of the  
superfluid component and the sound interact because of the nonlinear  
character of these equations. Explicit expressions for the scattered  
pressure and temperature are worked out in a first Born approximation,  
and care is exercised in delimiting the range of validity of the  
assumptions needed for this approximation to hold. An incident second  
sound wave will partly convert into first sound, and an incident first  
sound wave will partly convert into second sound. General considerations  
show that most incident first sound converts into second sound, but not  
the other way around. These considerations are validated using a vortex  
dipole as an explicitely worked out example. 
 \end{abstract} 
 
%\begin{flushleft} 
%PACS Numbers: 
%\end{flushleft} 
 
\newpage

%debut du texte

\section{Introduction}
The interaction of collective excitations with superfluid vortices has 
been a subject of intense research ever since the concept of mutual 
friction was introduced\cite{Gortermellink,HallVinen1,HallVinen2}. For 
reviews see\cite{Donnelly,Sonin}. Recent problems have 
included the question of the inertial mass of a vortex\cite{mass1,mass2}, 
the nature of the force on a moving vortex\cite{magnus}, possible 
generalizations to other systems that support vortices and collective 
excitations\cite{DemircanAoNiu}, as well as approaches based on the 
nonlinear Schr\"odinger equation\cite{nls1,nls2}.

As an alternative to an approach based on a microscopic description of a 
superfluid, it is possible to consider the well established equations of 
two fluid hydrodyna\-mics\cite{Khalatnikov}. These equations describe, 
among 
others, two modes of flow: One of them, obtained by linearizing the 
equations around a static, uniform state, is sound, both first and second. 
Another can be associated with superfluid vortical flow, in which the only 
dynamical quantity is the superfluid velocity. The corresponding 
vorticity is concentrated within thin filaments, whose circulation 
is quantized in units of $h/m$. This last fact comes into the two-fluid 
equations as an {\it ad-hoc} assumption.    Because of the nonlinearity 
of the hydrodynamic equations, these two modes interact, and as a result a 
sound wave that impinges on a superfluid vortex will be scattered. This 
paper is devoted to a study of this effect. Previous 
work\cite{LundSteinberg} studied the scattering of second sound waves by 
quantum vorticity very near the lambda point, at temperatures such that 
the superfluid density is very small compared to the normal fluid density. 
Earlier, Pitaevskii\cite{Pitaevskii} calculated the scattering of first 
sound by a quantum vortex in two dimensions in a Born approximation, and 
Fetter\cite{Fetter} in an exact form. Other nonlinear acoustic effects, 
related to the interaction of the sound waves with themselves, have also 
been studied\cite{Nemirovskii,Krysac}.

In this paper we study the scattering of sound (both first and second) by 
quantum vorticity within the framework of two fluid hydrodynamics, in a 
first Born approximation in three space dimensions. The reasoning follows 
earlier ideas that were developped in classical 
hydrodynamics\cite{LundRojas}, that have been succesfully tested in 
experiment\cite{bcp}. We hope that succesfully implementing the ideas of 
the present paper may lead to a new experimental probe to study the 
properties of quantum vorticity.

\section{Equations of the problem}
In this section we derive the equations needed to study the interaction 
of sound waves with quantum vorticity under the following assumptions: 
sound frequency is high compared with the inverse of all relevant time 
scales associated with the vortex motion,as well as with time scales 
associated with viscous and thermal 
relaxation effects; velocity amplitudes associated with sound are small 
compared to velocities associated with the vortices, and the latter are  
small compared to the speed of 
both first and second sound (even near the 
vortex cores); finally, thermal expansion effects are 
assumed to be negligible. Our analysis will be based in the 
phenomenological two-fluid model, in which Helium II, of density $\rho$, 
is regarded as a mixture of normal fluid, of density $\rho_n$, and a 
superfluid, of density $\rho_s$, with $\rho =\rho_s + \rho_n$. The flow of 
this system is described by normal and superfluid velocity fields $\Vn$ 
and $\Vs$. A characteristic feature of this model is that it predicts that 
small disturbances away from equilibrium values obey a wave equation, and 
that when the coefficient of thermal expansion is small enough, there are 
two modes of propagation: one (``first sound'') corresponds to density 
fluctuations as in normal fluids; the other (``second sound'') corresponds
to temperature fluctuations. The speed of propagation of second 
sound waves depends strongly on temperature and it vanishes at the Lambda 
point\cite{Donnelly}. The dynamics of the two-fluid model is described 
by the following set of nonlinear equations that quantify the conservation 
of mass, the conservation of linear momentum, the fact that the vorticity 
associated with the superfluid velocity is concentrated within vortex 
cores, and the fact that entropy is transported by the normal 
fluid only\cite{Khalatnikov}:

\begin{eqnarray}
{\partial \rho \over \partial t}  + \nabla \cdot \J & = & 0, \label{2F1}
 \\
{\partial J_i \over \partial t}  + \nabla_j \Pi_{ij} & = & 0,\label{2F2}
 \\ 
{\partial \Vs \over \partial t}  + (\Vs \cdot \nabla)\Vs & = & -\nabla \mu 
+ 
\f,\label{2F3}
 \\ 
{\partial (\rho S) \over \partial t}  + \nabla \cdot (\rho S \Vn ) & = &  
0,\label{2F4}
\end{eqnarray}
\noindent
where
\begin{eqnarray}
 \J & = & \rho_s \Vs + \rho_n \Vn,\label{D2F1}
 \\
 \Pi_{ij} & = & p \delta_{ij} + \rho_n V_{ni}V_{nj} + \rho_s 
V_{si}V_{sj},\label{D2F2}
 \\
 d\mu & = & {1\over \rho}d p - S d T - {\rho_n\over 2\rho}d 
W^2,\label{D2F3}
 \\
 \f & = & -{B\rho_n \over 2\rho}\bomega \times (\hat\omega_s \times \W) - 
{B'\rho_n 
\over 2\rho}\bomega \times \W\label{D2F4}
\end{eqnarray}\noindent
where  $p$, $S$ and $T$ are 
the  pressure, entropy per unit mass, and temperature, 
respectively, $\W \equiv \Vn - \Vs$, $\bomega \equiv \nabla \times \Vs$ 
is the superfluid vorticity, and 
$\hat\omega_s \equiv \bomega /|\bomega |$. $B$ and $B'$ are the usual 
mutual friction coefficients. Thermal and viscous diffusivity 
effects have been neglected assuming the sound frequency to be large. Eqn. 
(\ref{2F3}) contains terms that involve only the interaction of 
a superfluid vortex with itself (``self-induction term''); they  have been 
discarded because 
they do not couple to the sound waves. The entropy equation 
(\ref{2F4}) has terms cubic in the velocity; they have been omitted 
because they will be shown to be negligible below.

We describe the flow as the sum of fast {\it irrotational} ``sound''(\pv ) 
and 
slow {\it incompressible} ``vortical'' (\Us ) parts:
\begin{equation}
 {\bf V} = \pv + \Us, \quad \nabla \times \pv = 0, \quad \nabla \cdot \Us 
= 0, 
\quad v \ll U_s. \label{somme}
\end{equation}\noindent 
Note that the sound component $\pv$ contains both normal and superfluid  
components.
The vortical component \Us\ is a permanent superfluid flow, the  
equilibrium velocity field; it is a  
solution of the 
equations obtained by considering constant densities $\rho_{0n}$ and  
$\rho_{0s}$, temperature 
$T_0$ and entropy $S_0$, as no normal fluid flow is involved ($\Vn =  
0$), and from which $\nabla \cdot \Us = 0 $ follows as a consequence. 
This flow arises because vortex 
filaments move under their mutual and self-induction, possible external 
flows and appropriate boundary conditions.  The decomposition given in  
Eq. (\ref{somme}) is unique when boundary conditions for $\Us$ are  
specified. For instance, that it vanish sufficiently fast at infinity so  
that a scattering formulation makes sense.  
It is simple to check that under these conditions the dynamics is  
completely governed by the equation 
\beq 
{\partial \Vs \over \partial t}  + (\Vs \cdot \nabla)\Vs =  
-{1\over \rho_0} \nabla p + {\rho_{0n}\over 2\rho_0} \nabla \Vs^2 +\f_0 
\label{eq:incomp} 
\eeq 
where $\f_0$ is the mutual friction force $\f$ evaluated at constant 
densities. Taking the curl  
of Eq. (\ref{eq:incomp}) yields an evolution equation for the superfluid  
vorticity, from which the (incompressible) superfluid velocity may be  
deduced. Taking the divergence of Eq. (\ref{eq:incomp}) yields an  
elliptic equation for the pressure in terms of the superfluid velocity.  
Thus, just as in classical  
incompressible hydrodynamics, the pressure is not a dynamical variable,  
but it is determined at each instant of time by the current value of the  
superfluid velocity. 
 
The flow \pv\ includes both first and 
second 
sound, and consists of small deviations of density $\rho'$, temperature 
$T'$, 
entropy $S'$ and pressure $p'$ away from their equilibrium values (note 
that, 
due to the flow \Us , the equilibrium value $p_0$ of the pressure is not 
a constant). The interaction between the two modes, sound and vortical, 
 is obtained by linearizing the 
equations around the vortical solution \Us . The result on equation 
(\ref{2F1}) is

\begin{equation}
{\partial \rho' \over \partial t}  + \rho_s^0\nabla \cdot \vs + 
\rho_n^0\nabla 
\cdot \vn= -(\Us \cdot \nabla)\rho'_s, \label{dev1}
\end{equation}
\noindent
and we note that the right hand member is linear in \Us . 
Applying the 
same procedure on (\ref{2F2}) leads to

\begin{equation}
\rho_s^0{\partial v_{si} \over \partial t} + \rho_n^0{\partial v_{ni} 
\over 
\partial t} + \nabla_i p'= -{\partial (\rho'_s U_{si})\over \partial t} - 
\nabla_j\left(\rho'_s U_{si}U_{sj} + \rho_s^0 U_{sj}v_{si} + \rho_s^0 
U_{si}v_{sj}\right),\label{dev2}
\end{equation}
\noindent
in wich a quadratic term, $\nabla_j(\rho'_s U_{si}U_{sj})$, appears. Let 
us 
compare this term with $({\partial (\rho'_s U_{si})/ \partial t})$ :
\begin{eqnarray}
{\partial (\rho'_s U_{si})\over \partial t} & = & 
\underbrace{U_{si}{\partial 
\rho'_s \over \partial t}}_{\simeq \nu_0 \rho'_s U_s} + 
\underbrace{\rho'_s{\partial U_{si}\over \partial t}}_{\simeq (1/\Delta  
t)\rho'_s 
U_s}
\label{approxdev2a} \\  
 & &  \nonumber \\ 
\nabla_j(\rho'_s U_{si}U_{sj}) & = &  
\underbrace{U_{si}U_{sj}\nabla_j \rho'_s}_{\simeq k_0 \rho'_s (U_s)^2} + 
\underbrace{\rho'_s U_{sj} \nabla_j U_{si}}_{\simeq (1/L) \rho'_s 
(U_s)^2}\label{approxdev2b}
\end{eqnarray}
\noindent
where $\nu_0$ is the wave frequency, $k_0$ its wavenumber, $\Delta t$ a 
typical 
time scale  
for the vortical flow, and $L$ a typical length scale. We assume $\nu_0 
\gg 
1/\Delta t$, 
so that the second term in the r.h.s. of \refpar{approxdev2a} is 
negligible; 
the wave probes the vortical flow on typical scales equal to its 
wavelength, 
so that $k_0 \simeq 1/L$, and
\begin{equation}
{|\nabla_j(\rho'_s U_{si}U_{sj})| \over |({\partial (\rho'_s U_{si})/ 
\partial 
t}) |} \simeq {U_s \over c_i} \ll 1\label{approxdev2c}
\end{equation}
\noindent
where $c_i$ is either the first ($i = 1$) or the second ($i = 2$) sound 
velocity. However, the last inequality is not obviously verified, because 
the velocity $U_s$ may take large values near the vortex core. Sound  
scattering effects, as most wave mechanics effects, are most dramatic 
when interference effects are dominant. This occurs when the wavelength is 
comparable to the length scale of the structure responsible for the 
scattering. Consequently, sound scattering will ``see'' structures with a 
typical size equal to  the wavelength of the sound 
wave, say $\lambda_i$. At this distance from the core of a vortex line, 
$U_s 
\simeq \hbar /(m_{He} \lambda_i)$, where $m_{He}$ is the mass of an helium 
atom. Thus if $\nu_0$ is the frequency of the wave, we have

\begin{equation}
{U_s \over c_i} 
\ll 1 \Longleftrightarrow \nu_0 \ll {m_{He} c_i^2 \over \hbar} \,\approx 
\,
\cases{ 5\, 10^{12} \,Hz \; \hbox{(first sound)} \cr
 8\, 10^{9} \,Hz \; \hbox{(second sound)} \cr}. \label{approxdev2d}
\end{equation}
\noindent
The condition is easy to realize for first sound waves, and clearly 
compatible 
with our previous requirement that the time scale of the sound wave should 
be 
much smaller than that of thermal and viscous diffusion effects. For the 
second 
sound wave, the numerical value is calculated for $c_2 \approx 10\, m/s$, 
which 
is the minimum value of the second sound velocity between $1.75\,K$ and 
$2.15\,K$ ($c_2 = 0$ at \tc, but it is only very near the $\lambda$-point 
that 
it becomes less than $10\,m/s$). Once again, the required inequality is 
easy to 
fulfill, so that it is consistent to neglect 
the term which is quadratic in $U_s$ in (\ref{dev2}) whatever the nature 
of 
the incident wave. This term however, will be responsible for the  
spontaneous generation of sound, a topic that is outside the scope of  
the present work (see \cite{vortexsound} for the situation in classical  
hydrodynamics).

\bigskip

Let us now turn to the linearization of (\ref{2F3}).  Two new terms 
appear.  The first one reads $(\rho'/\rho_0)\nabla p_0$; but from the 
Bernouilli formula we know that $\nabla p_0 \simeq \rho_0 U_s^2$.  This 
term vanishes in the case of a second sound wave, because in this case 
\cite{LandauLifshitz} $\rho' = 0$, and for a first sound wave it is of 
order $U_s/c_1$ in comparison of linear terms in $U_s$, and may thus be 
neglected. 

The other term comes from the linearization of the mutual friction 
force $\f$, and we have to calculate $(\rho'_n U_s^2) / (\rho_n^0 
U_{s}v_{s})$.  In the case of a first sound wave, this ratio is 
obviously of order $U_s/c_1$, thus negligible under the same 
approximations as before.  For a second sound wave, one needs to be 
more careful.  In that case \cite{LandauLifshitz}, $\rho' \approx 0$, 
but $\rho'_s = (\partial \rho_s /\partial T)_0 T' \neq 0$, and $T' 
\simeq -(c_2/S_0)v_s$ where $c_2$ is the velocity of second sound 
waves.  It is possible to show that the quadratic terms are negligible 
at sufficiently high temperatures [however, not too close to the 
$\lambda$-point; see the discussion following 
(\ref{approxdev2d})] ; in this limit, the thermodynamic quantities are 
accurately given by the roton part of the excitation spectrum 
\cite{Khalatnikov}.  We get 

\begin{equation}
{\rho'_n U_s^2 \over \rho_n^0 U_{s}v_{s}} = \left({\rho_0 \over 
\rho_n^0}\right)\left({\Delta_r/T_0 - 1/2\over \Delta_r/T_0 + 
3/2}\right)\left({c_2 
U_s \over 3(k_B T_0/\pi_0)^2}\right),\label{approx2s}
\end{equation}
\noindent
where $k_B$ is the Boltzmann's constant, $\Delta_r$ is the energy of the 
rotons 
and $\pi_0$ their impulsion. As soon as $T_0 \geq 1.75\,K$, the ratio in 
the 
r.h.s. of (\ref{approx2s}) is less than $1\%$. To obtain this estimate, we 
assumed $U_s \simeq 10^{-2} c_2$ as before, so that the quadratic terms 
are 
all negligible, 
with the same validity range as before. The equation derived from 
(\ref{2F3}) 
thus reads

\begin{eqnarray}
& &{\partial \Vs \over \partial t}  + {1\over \rho_0}\nabla p' - S_0 
\nabla T' 
= - {\rho_s^0 \over \rho_0}\left[(\vs \cdot \nabla)\Us + (\Us \cdot 
\nabla)\vs\right] \nonumber \\ &  & \quad -{\rho_n^0 \over 
\rho_0}\left[(\vn 
\cdot \nabla)\Us + (\Us \cdot \nabla)\vn\right] -\left[1 - {B'\over 
2}\right]{\rho_n^0 \over \rho_0}(\vn -\vs)\times \bomega  \label{dev3}\\ & 
 & 
\qquad \qquad 
-{B \rho_n^0 \over 2\rho_0}\bomega \times \left[\hat\omega_s \times (\vn - 
\vs)\right]
. \nonumber  
\end{eqnarray}
\noindent
As for the last equation, the linearization of (\ref{2F4}),  it doesn't 
contain a source term, and reads

\begin{equation}
S_0{\partial \rho'  \over \partial t} + \rho_0{\partial S'  \over \partial 
t} + 
\rho_0 S_0 \nabla \cdot \vn  = 0.\label{dev4}
\end{equation}
\noindent 
The absence of source terms in this equation is due to the neglect of 
diffusive effects, 
and 
moreover comes from the fact that the mutual friction terms are all 
cubic in 
the velocities, hence quadratic in \Us, so that we can neglect them under 
the same assumptions that led to 
equations \refpar{dev2} and \refpar{dev3}.

In liquid helium, the coefficient of thermal expansion $\beta = 
-(1/\rho)\partial \rho /\partial T$ is extremely small, and may be 
neglected. 
It simplifies greatly the equation of sound propagation, because in this 
limit the free propagation of first and second sound are completely 
decoupled. Indeed, after some algebra, using the fact that the sound wave 
flow is irrotational and that the vortical flow is divergence-free, one 
finds for the equations of propagation

\begin{equation}
\left({1\over c_1^2}  {\partial^2 \over \partial t^2}  - \Delta 
\right)p'  = 
F_1,
\label{son1} 
\end{equation}
where
\[
F_1 \equiv \rho_s^0\biggl\{\Delta (\Us \cdot \vs) + \nabla \cdot 
\bigl[\bomega \times \vs + \Us (\nabla \cdot \vs)\bigr]\biggr\}, 
\]

\noindent
and 

\begin{equation}
\left({1\over c_2^2}{\partial^2 \over \partial t^2} - \Delta \right)T' 
= 
F_2, 
\label{son2} \\ 
\end{equation}
where
\bea
F_2 \equiv {1 \over 
\rho_0S_0}\left\{(\Us \cdot \nabla )\left({\rho_0 \over 
\rho_s^0}\,{\partial 
\rho'_s \over \partial t} + \rho_s^0 \nabla \cdot \vs \right)\right.  -  
\rho_n^0 \bigl[\Delta (\Us \cdot \vn) + \nabla \cdot (\bomega \times 
\vn)\bigr] & & \nonumber \\
 \qquad - \left[1 - {B'\over 
2}\right]\rho_n^0\nabla \cdot \bigl[(\vn -\vs)\times \bomega\bigr] 
 -  \left. {B \rho_n^0 \over 2}\nabla \cdot \bigl[\bomega \times 
\bigl(\hat\omega_s \times (\vn - 
\vs)\bigr)\bigr]\right\} & & \nonumber .
\eea
Equations (\ref{son1}) and (\ref{son2}) are accurate to first order in 
$v_s  / U_s  $, $U_s / c_{\rm i}$ (i=1,2), and $\nu_0 \delta t$, in the 
temperature range $[1.75, 2.15]K$. To 
zeroth order ($F_1 = F_2 = 0$) they respectively describe the usual 
equations for 
free first and second sound propagation.

\section{Scattering of sound waves by vorticity}

Equations (\ref{son1}) and (\ref{son2}) can both be written as integral 
scattering equations, 
\begin{equation}
SOUND_{scatt.} = SOUND_{inc.} + G_i \ast F_i \label{convolution}
\end{equation}\noindent
where $SOUND$ is either a pressure wave, in the case of first sound, or a 
temperature 
wave, in the case of second sound; $i = 1,2$ corresponding to first or 
second sound: $G_i$ is the retarded Green function for the 
wave 
equation, with vanishing boundary conditions at infinity,
\begin{equation}
G_i = G_i(\x - \x',t-t') = (4\pi |\x -\x'|)^{-1}\, \delta 
(t-t'-c_i^{-1}|\x 
-\x'|), \label{Green}
\end{equation}\noindent
and the star $*$ denotes a space-time convolution. The scattering problem 
is ready to be solved using the Born approximation, 
which 
consists of injecting in $F_1$ and $F_2$ the values of $\rho'_n$, \vs\ and 
\vn\ 
given by the incident (first or 
second) sound wave. 
In either case, both  $F_1$ and $F_2$ are different from 0, so that a 
second sound wave 
illuminating a vortex leads to a scattered {\it first} sound wave, and 
conversely. 

\subsection{Scattered second sound from incident second sound}

We consider the scattering of an incident plane temperature wave.
\begin{equation}
T' = T'_i \cos (k_{02}  x - \nu_0 t)  = {T'_i \over 2} e^{i(k_{02}  x - 
\nu_0 
t)} + CC, \label{Tinc2}
\end{equation}
\noindent
where $CC$ stands for "complex conjugate", and $k_{02} \equiv \nu_0/c_2$ 
is the 
relevant wave vector for a {\it second} sound wave of frequency $\nu_0$. 
It is 
easy to show \cite{LandauLifshitz} that 
the velocity field is given by

\begin{equation}
\vs = \vi e^{i(k_{02}  x - \nu_0 t)} + CC,\quad \vn = -{\rho_s^0 \over 
\rho_n^0}\vs,\quad \vi = -{S_0 T'_i \over 2 c_2} \vecinc, \label{Vinc2}
\end{equation}
\noindent
where $\vecinc$ is a unit vector in the incident direction. Moreover, we 
have

\begin{equation}
\rho'_s = \left({\partial \rho_s \over \partial T}\right)^0_{\rho} T' + 
\left({\partial 
\rho_s \over \partial \rho}\right)^0_T \rho' \approx  \left({\partial 
\rho_s 
\over 
\partial T}\right)^0_{\rho} T'\label{rhoinc2}
\end{equation}
\noindent
where the superscript $"0"$ indicates that all partial derivatives are 
taken at equilibrium, and the last approximation is justified by the 
smallness of the 
coefficient 
of 
thermal expansion $\beta$ (\cite{LandauLifshitz}). It is thus possible to 
show that $F_2 
= 
- \nabla \cdot {\bf D}_2$, where the vector ${\bf D}_2$ is given by

\begin{eqnarray}
 {\bf D}_2 &=& \underbrace{-{\left({\partial \rho_s / \partial 
T}\right)^0_{\rho} 
\over 
\rho_s^0S_0}{\partial \rho'_s \over \partial t}\Us}_{\equiv (a)} \; + \;
\underbrace{{\rho_n^0 \over \rho_0S_0} \left[\nabla (\Us\cdot \vn ) + 
\bomega 
\times \vn + \Us(\nabla\cdot\vn)\right]}_{\equiv (b)} \nonumber \\ & & 
\qquad 
+ \underbrace{\left[1 - {B'\over 2}\right]{1 \over S_0} \vs\times 
\bomega}_{\equiv (c)}\quad +
 \underbrace{ {B  \over 2 S_0}\;\bomega \times \left(\hat\omega_s \times 
\vs\right)}_{\equiv (d)} \label{F22}
\end{eqnarray}

Let $T'_{(a)}$ be the contribution of the term $(a)$; the detailed 
calculation 
of this term uses the following steps:

\begin{eqnarray}
{T'_{(a)}\over T'_i} &=& {\left({\partial \rho_s \over \partial 
T}\right)^0_{\rho} 
\over 
2\rho_s^0S_0} \int \!\!\!\diff t'\!\!\int \!\!\!\diff^3x' G_2(\x - 
\x',t-t') 
\nabla \cdot 
\left[\Us (\x',t')e^{-i\nu_0 t'}(-i\nu_0)e^{i \kzerodeux \cdot 
\x'}\right] \nonumber \\ &=& { c_2\left({\partial \rho_s \over \partial 
T}\right)^0_{\rho} \over 2\rho_s^0S_0}(-ik_{02})\int \!\!\!\diff^3x' 
\widetilde{G}_2(\x 
- 
\x',\nu) \nabla \cdot \left(\wus(\x',\nu - \nu_0)e^{i \kzerodeux \cdot 
\x'}\right)\nonumber \\ &=& { c_2\left({\partial \rho_s \over \partial 
T}\right)^0_{\rho} \over 2\rho_s^0S_0}\,{(-ik_{02})\over 2}\left[\int 
\!\!\!\diff^3x' 
\widetilde{G}_2\nabla \cdot \left(\wus e^{i \kzerodeux \cdot \x'}\right) - 
\int \!\!\!\diff^3x' \nabla\widetilde{G}_2 \cdot \wus e^{i \kzerodeux 
\cdot 
\x'}\right]\nonumber \\ &=& {T_0 S_0\left({\partial \rho_s \over \partial 
T}\right)^0_{\rho} \over C\rho_n^0}\;{i\nu\pi^2 e^{i\nu r /c_2}\over r 
c_2^2}\;{1 
\over 2(1 - \cos \theta )}\;(\vecinc\times \hat r)\cdot 
\wbomega(\q_{2\to 2},\nu-\nu_0)\label{T21} 
\end{eqnarray}
\noindent
where 

\begin{equation}
\wbomega(\q,\nu) = {1\over (2\pi)^4}\int \!\!\diff t\diff^3x e^{i(\nu 
t - \q\cdot\x)}\bomega (\x,t) \label{TFom}
\end{equation}
\noindent
is the space-time Fourier transform of the vorticity, and 

\begin{equation}
\q_{2\to 2} \equiv {\nu \over c_2}\hat r - {\nu_0 \over c_2}\vecinc 
\label{defq2par2}
\end{equation}
\noindent
is the momentum transfer, with $\hat r$ a unit vector in the direction of 
observation, and \vecinc\ a unit vector in the direction of propagation of 
the 
incident wave. In the last line of (\ref{T21}), we introduced the 
scattering angle $\theta$, the heat capacity $C$, and we also used the 
asymptotic 
behavior at large distance of $\widetilde G_i$, which we quote here for 
reference 

\begin{equation}
\cases{\widetilde G_i \stackrel{x \gg x'}{\simeq} ({e^{i\nu r /c_i}/ 4 \pi 
r})e^{i\nu 
\hat r \cdot \x' /c_i}, 
\cr  \cr 
\nabla \widetilde G_i \stackrel{x \gg 
x'}{\simeq} 
({i \nu / c_i})({e^{i\nu r /c_i}/ 4 \pi r})e^{i\nu \hat r \cdot \x' 
/c_i}(-\hat r) 
\cr \cr 
\Delta \widetilde G_i \stackrel{x \gg x'}{\simeq} 
(i\nu/c_i)^2({e^{i\nu r /c_i}/ 4 \pi r})e^{i\nu 
\hat r \cdot \x' /c_i}}\label{asympG}
\end{equation}
\noindent
Moreover, we used the fact that the difference between the frequencies of 
the 
incident and scattered wave is of the order of the typical frequency of 
$\Us 
(\x,t)$, thus much smaller than $\nu_0$, so that $\nu/c_2 \approx 
\nu_0/c_2$. 
Note that this assumption is very easy to fulfill, if one consideres 
\refpar{approxdev2d}.

The source terms are linear in \Us, which is completely characterized by 
its 
vorticity, being a divergenceless field; the scattered field should thus 
be 
expressed as a function of the vorticity, and more precisely of its 
Fourier 
transform. The dependence of $T'_{(a)}$ on $(\vecinc\times \hat r)\cdot 
\wbomega $ was to be expected, because it is the only possible scalar 
expression depending both on the direction of propagation of the incident 
wave 
$\vecinc$, on the direction of observation $\hat r$ and on the vorticity. 
The 
only term in (\ref{F22}), leading to a different behavior, is (d) which is 
second order in vorticity. The calculation of the contributions of the 
remaining terms in (\ref{F22}) requires only integrations by parts, and 
leads 
to the following result :

\begin{equation}
{T'_{2\to 2}(\x,\nu) \over T'_i} = {i\nu\pi^2 e^{i\nu r /c_2}\over r 
c_2^2}\;\left[H_c(\theta)(\vecinc\times \hat r)\cdot 
\wbomega - {B\over 2}\hat r_i \vecinc_j 
\widetilde{\Omega}_{ij}\right]\label{T2par2}
\end{equation}\noindent
where the Fourier transforms are both taken at the point $(\q_{2 
\to 2},\nu-\nu_0)$ (see \refpar{defq2par2}), and 
where 
\begin{equation}
 H_c(\theta) \equiv \left(1 - {B'\over 2}\right) - {1 \over 2(\cos \theta 
-1)}\,{T_0 S_0 \over C\rho_n^0}\,\left({\partial \rho_s \over \partial 
T}\right)^0_{\rho} - {\rho_s^0 \over \rho_0}\,{\cos \theta \over \cos 
\theta -1}\label{defHc} 
\end{equation}\noindent
and 
\begin{equation}
 \Omega_{ij} \equiv {\omega_i\omega_j\over \omega} - 
\omega\delta_{ij}.\label{defOmega} 
\end{equation}\noindent
The terms that are multiplied by $B$ and $(1 - B'/2)$ have already been 
calculated 
by 
Lund \& Steinberg \cite{LundSteinberg}, but in their approximation scheme 
there should be 
a factor 
of $\rho_n^0/\rho_0$, which they dropped because in their approximations 
it is 
very close to $1$. Our calculation shows that term to be an artifact of 
the 
approximations used, and it dissapears in our more accurate scheme. 
The second term in the r.h.s. of \refpar{defHc} was 
also 
obtained in \cite{LundSteinberg}. The last one is a correction of their 
result, when 
$\rho_s$ is not negligible; it is exactly the same term (up to the factor 
$\rho_s^0/\rho_0$) that is found for the scattering of 
ordinary 
sound by a vortex, in a classical fluid \cite{LundRojas}. This was to be 
expected 
since the 
corresponding source term comes from the terms quadratic in velocity of 
the 
conservation of momentum equation. 

\subsection{Scattered first sound from incident second sound}

A second sound temperature wave, as we have seen, couples to the 
superfluid vortical velocity field; it causes vibrations of the vortices, 
and 
those vibrations generate a scattered second sound wave, but also a {\it 
first} sound wave : The vortical velocity field couples the two differents 
acoustic modes in Helium II, even in our approximation, $\beta = 0$, in 
which 
their respective equations of free propagation are completely decoupled.

The source $F_1$ in \refpar{son1} is a sum of three different 
contributions, 
and they finally have to be expressed as a function of $(\vecinc\times 
\hat 
r)\cdot 
\wbomega $ only, as we explained before. This is obvious for the term 
$\nabla 
\cdot (\bomega \times \vs )$, but not for the two other terms. We may 
proceed 
in the following way to show that it is indeed the case. After a  
Fourier 
transformation in time, we have to calculate

\begin{eqnarray}
& & \int \!\!\!\diff^3x' \widetilde{G}_2 \left[ \Delta(\wus \cdot \vs) + 
\nabla 
\cdot \left(\wus(\nabla \cdot \vs)\right)  \right] = \label{Scatt1par2-1} 
\\ & 
& 
  \int \!\!\!\diff^3x' \left[ \Delta\widetilde{G}_2 (\wus \cdot \vs)  - 
\alpha \nabla \widetilde{G}_2\cdot \left(\wus(\nabla \cdot \vs)\right) + 
(1 - 
\alpha)\widetilde{G}_2( \wus \cdot \nabla)(\nabla \cdot \vs) 
\right]\nonumber 
\end{eqnarray}
\noindent
where $\alpha$ is, for the time being, an unknown parameter. We have 
integrated 
by parts the integral containing the Laplacian operator, and we have split 
the second 
one, using integration by parts for the term in factor of $\alpha$, and 
$\nabla 
\cdot \Us = 0$ in order to simplify the last term. The r.h.s. of 
\refpar{Scatt1par2-1} takes the form

\begin{equation}
{e^{i\nu r /c_1}\over 4 \pi r}\,{i\nu \over c_1}\,\int \!\!\diff^3 x' 
X(\x',\nu - \nu_0)e^{-i\q_{2\to 1} \cdot \x'} \label{Scatt1par2-2} 
\end{equation}\noindent
where 
\begin{equation}
\q_{2\to 1} \equiv {\nu\over c_1}\,\hat r - {\nu_0 \over c_2}\,\vecinc, 
\label{defq1par2}
\end{equation}
and $X$, which is a function of $(\x', \nu 
- \nu_0)$ like \wus, is given by
\begin{equation}
X \equiv i\left[k_1(\wus \cdot \vi) + \alpha (\hat r \cdot \wus)(\vi \cdot 
\kzerodeux) + {1 - \alpha \over k_1}(\kzerodeux \cdot \wus)(\vi \cdot 
\kzerodeux)\right] 
\label{Scatt1par2-3} 
\end{equation}\noindent
where $k_1 \equiv \nu/c_1$. The difference between the frequencies of the 
incident and scattered wave is of the order of the typical frequency of 
$\Us 
(\x,t)$, thus much smaller than $\nu_0$, so that $k_1 \approx \nu_0/c_1$. 
However,  $k_{02} = \nu_0/c_2$ and the sound velocities of the two 
acoustic 
modes 
are very different, so that we do not have the simplifying assumption 
$k_{02} 
\approx k_1$, as in the scattering of sound by vorticity in simple fluids 
\cite{LundRojas}, 
or as in 
the previous case of second sound scattering from incident second sound. 
But, introducing an unknown coefficient $\alpha$ allows 
us 
to {\it choose} it in order that $X$ be proportional to $(\vecinc\times 
\hat 
r)\cdot 
(\q_{2\to 1}\times\wbomega) $; indeed, expressing the equation

\begin{equation}
X(\x', \nu - \nu_0) = {\cal C} (\vecinc\times \hat r)\cdot 
\left[\q_{2\to 1} \times\wbomega(\x', \nu - 
\nu_0)\right],\label{proportion} 
\end{equation}
\noindent
with ${\cal C}$ an (as yet) unknown coefficient of proportionality, 
in terms of 
the components of the relevant vectors in the scattering plane,

\begin{equation}
\kzerodeux = \left( \begin{array}{c} k_{02} \\ 0 \\ 0 
\end{array}\right),\;
{\bf k}_1 = \left( \begin{array}{c} k_1\cos \theta \\ k_1\sin \theta \\ 0 
\end{array}\right),\; \hat r = \left( \begin{array}{c} \cos \theta \\ \sin 
\theta \\ 0 \end{array}\right),\;\vi = \left( \begin{array}{c} v^i \\ 0 \\ 
0 
\end{array}\right),\;\wus = \left( \begin{array}{c} \tilde U^x_s \\ \tilde 
U^y_s 
\\ 
\tilde U^z_s \end{array}\right) ,
\label{vecteurs} 
\end{equation}
\noindent
the elimination of $\tilde U^x_s$ and $\tilde U^y_s$ leads to a linear 
system of two equations, for the two unknowns $\alpha$ and $\cal C$, whose 
solution is

\begin{equation}
{\cal C} = {1 + (c_2/c_1) \over 2(c_2/c_1)\cos \theta -1 -(c_2/c_1)^2} 
.\label{coeffprop} 
\end{equation}
\noindent
Note that if we let $c_1 = c_2$, we recover the expression valid in 
simple fluids with only one kind of acoustic wave \cite{LundRojas}. Using 
\refpar{coeffprop}, we get the pressure field 
scattered by a vortical region illuminated by second sound,

\begin{equation}
{\tilde p_{2\to 1}'(\x,\nu) \over \rho_s^0S_0T'_i} = - \,{i\nu\pi^2 
e^{{i\nu r 
/ c_1}}\over r c_2c_1}\,\Pi(\theta)(\vecinc\times \hat r)\cdot 
\wbomega(\q_{2\to 1}, \nu - \nu_0), \label{p1par2} 
\end{equation}
\noindent
where $\q_{2\to 1}$ is defined by \refpar{defq1par2} and $\Pi(\theta)$ is 
given by
\begin{equation}
\Pi(\theta) \equiv  
{2\left(c_2 / c_1\right)\cos 
\theta \over 2\left(c_2 / c_1\right)\cos \theta -1 
-\left(c_2 / c_1\right)^2}. \label{defPi}
\end{equation}
The reader is reminded
that this equation, together with \refpar{T2par2}, is valid for 
temperatures in 
the range $[1.75,2.15]\,K$, under the general assumptions of section {\bf 
2}.

\subsection{Scattered first sound from incident first sound}

We consider now the scattering of an incident plane pressure wave.

\begin{equation}
p' = p'_i \cos (k_{01}  x - \nu_0 t)  = {p'_i \over 2} e^{i(k_{01}  x - 
\nu_0 
t)} + CC, \label{pinc1}
\end{equation}
\noindent
where $CC$ stands for "complex conjugate", and now $k_{01}$ is the 
relevant 
wave 
vector for {\it first} sound waves,  $k_{01}\equiv \nu_0/c_1$, in 
contrast with sections {\bf 3.1} and {\bf 3.2}. It is easy to show 
\cite{LandauLifshitz} 
that 
the velocity field is given by

\begin{equation}
\vs \approx \vi e^{i(k_{01}  x - \nu_0 t)} + CC,\quad \vn = \vs,\quad \vi 
= { 
p'_i \over 2 \rho_0 c_1} \vecinc, \label{Vinc1}
\end{equation}
\noindent
Moreover, we have

\begin{equation}
\rho' = {p'_i \over 2 c_1^2}\,e^{i(k_{01}  x - \nu_0 t)} + CC 
\label{rhoinc1}
\end{equation}
\noindent
where the contribution of terms proportional to the coefficient of 
thermal expansion $\beta$ has been neglected.

First sound scattering from the vortical flow is given by the source term 
in 
the 
equation of propagation for the pressure, \refpar{son1}. This term, apart 
from 
a 
factor $\rho_s^0$ instead of $\rho_0$, is identical to the corresponding 
one in 
normal fluids \cite{LundRojas}, hence the scattered pressure field is 
exactly the 
same as for ordinary sound scattering by vorticity in normal fluids, and 
reads

\begin{equation}
{\tilde p_{1\to 1}'(\x,\nu) \over p'_i} = {\rho_s^0 \over \rho_0} 
\,{i\nu\pi^2 
e^{{i\nu r / c_1}}\over r c_1^2}\,{\cos \theta \over \cos \theta -1 
}(\vecinc\times \hat r)\cdot 
\wbomega(\q_{1\to 1}, \nu - \nu_0).\label{p1par1} 
\end{equation}
\noindent
where 
\begin{equation}
\q_{1 \to 1} \equiv {\nu\over c_1}\hat r - {\nu_0\over c_1}\vecinc; 
\label{defq1par1}
\end{equation}
Note that the approximation 
 $\nu/c_1 \approx k_{01}$ is valid here, as in ordinary fluids. This 
ensures 
the same scattering of (ordinary) sound waves in both cases.

\subsection{Scattered second sound from incident first sound}

Let us consider now the second sound scattered by a vortical flow 
illuminated 
by a first sound wave. In a first sound wave, neglecting the coefficient 
of thermal expansion $\beta$ \cite{LandauLifshitz}, we have $\vs = \vn$, 
so that the 
mutual friction  terms do not contribute to the scattering. Moreover, 
equation 
\refpar{rhoinc2} has to be replaced by
 
\begin{equation}
\rho'_s = \left({\partial \rho_s \over \partial T}\right)^0_{\rho} T' + 
\left({\partial 
\rho_s \over \partial \rho}\right)^0_T \rho' \approx  \left({\partial 
\rho_s \over \partial \rho}\right)^0_T \rho' \label{rhosinc1}
\end{equation}
\noindent
because $T'\approx 0$ for a first sound wave. The relevant expression of 
$F_2$ in (\ref{son2})  
now reads $F_2 = - \nabla \cdot {\bf D}'_2$, with ${\bf D}'_2$ given by

\begin{equation}
 {\bf D}'_2 = {1\over \rho_0 S_0}\,\left[1-{\rho_0 \over 
\rho_s^0}\left({\partial 
\rho_s \over \partial \rho}\right)^0_T\right]{\partial \rho \over \partial 
t}\Us 
+ {\rho_n^0 \over \rho_0 S_0}  \biggl[\nabla (\Us\cdot \vn ) + \bomega 
\times \vn + \Us(\nabla\cdot\vn)\biggr]  \label{F21}
\end{equation}
\noindent
The calculation now proceeds just as in section {\bf 3.2}; the wave vector 
coming from the Green function $G_2$, ${\bf k}_2 \equiv (\nu/c_2)\hat r$, 
is 
very different in amplitude from the incident wave vector, $\kzeroun = 
(\nu_0/c_1)\vecinc$, even if $\nu \approx \nu_0$, because the sound 
velocities 
$c_1$ and $c_2$ are  very different. We thus have to use the same trick as 
in 
equation \refpar{Scatt1par2-1}, and the result reads

\begin{equation}
\widetilde{T}'_{1\to2}(\x,\nu) = p'_i \Theta(\theta){i\nu\pi^2 e^{{i\nu r 
/ c_2}}\over r c_2c_1}(\vecinc\times \hat r)\cdot 
\wbomega(\q_{1\to2}, \nu - \nu_0).\label{T2par1}
\end{equation}
\noindent
where $\theta$ is the scattering angle, the scattering wavevector 
$\q_{1\to2}$ is

\begin{equation}
\q_{1\to2} \equiv {\nu \over c_2}\hat r - {\nu_0 \over c_1}\vecinc  
,\label{defq2par1}
\end{equation}
\noindent
and $\Theta(\theta)$ is given by

\begin{eqnarray}
\Theta(\theta) &\equiv &{(c_2 / c_1) \over 1 + (c_2 / c_1)^2 - 2(c_2 / 
c_1)\cos\theta} \times \nonumber \\ & &\quad \times \left\{{1 \over 
\rho_0c_1c_2}\,{S_0\rho_s^0 \over C \rho_n^0}\left[{\rho_0 \over 
\rho_s^0}\left({\partial 
\rho_s \over \partial \rho}\right)^0_T -1\right]T_0 + 
{2\rho_n^0 \over \rho_0^2 S_0}\cos \theta\right\}.\label{coeffT2par1}
\end{eqnarray}
The reader is reminded 
that this equation, together with \refpar{p1par1}, is valid for all 
temperatures, under the general assumptions of section {\bf 2}.

\section{Estimates and example}

In this section we estimate the relative energy scattered in each 
acoustic mode in the case of an incident first and second sound wave. 
Then we choose as an example target a rectilinear vortex dipole. The 
simplest 
possible flow is that of a rectilinear vortex; however in this case, 
the scattered waves \refpar{T2par2} and \refpar{p1par1} diverge in 
the forward scattering direction $\theta \to 0$, and this divergence 
disappears with a vortex dipole \cite{berthetlund}.

The time-averaged energy density of a sound wave is 
\bea
\overline{E} = \rho_s^0 \overline{v_s^2} + \rho_n^0 \overline{v_n^2},
\label{enerdens}
\eea 
where the overline denotes a time average over one period of 
oscillation \cite{Khalatnikov}. Far away from the target, the scattered 
wave may be approximated by a plane wave, so that the energy flow is given 
by $I = \overline{E}c$, where $c$ is the velocity of either a first or 
second sound wave. To calculate the superfluid and normal velocity fields, 
we use the following set of equations 
\bea
 \rho_s^0 {\partial \Vs \over \partial t} + \rho_n^0 {\partial\Vn \over 
 \partial t}  + \nabla p' &=& 0,\label{lin1}\\ 
 \rho_n^0 {\partial \over \partial t}( \Vn  - \Vs)  + \rho_0 S_0 \nabla T' 
 &=& 0.\label{lin2}
\eea
They are obtained by linearizing and combining \refpar{2F2} and 
\refpar{D2F1} in the first case, and \refpar{D2F3}, \refpar{2F3} and 
\refpar{2F2} in the second case. The energy flow for each kind of wave 
thus reads
\bea
I_{1} &=& {1\over \rho_0}\;\overline{|\nabla p'|^2}\;{c_1 \over 
\nu^2}\quad 
\hbox{ (first sound)} \label{enerflow1} \\ I_{2} &=& 
{\rho_s^0 \rho_0 S_0^2\over \rho_n^0}\;\overline{|\nabla T'|^2}\;{c_2 
\over \nu^2} \quad \hbox{ (second sound)} \label{enerflow2}
\eea
The gradients in \refpar{enerflow1} and \refpar{enerflow2} must 
be evaluated at leading order in $1/r$ for consistency.

Define $\wbomegap$, the component of superfluid vorticity perpendicular 
to the scattering plane, through the relation
\begin{equation}
\label{vortperp} 
\sin \theta \quad \wbomegap(\q_{{\rm i} \to {\rm j}}) \equiv
(\vecinc\times \hat r)\cdot 
\wbomega(\q_{{\rm i} \to {\rm j}})
\end{equation}
and, for the sake of argument, consider the following estimates:
\bea
 H_c(\theta) & \sim & 1 \nonumber \\
\Pi(\theta) & \sim & c_2 / c_1 \nonumber \\
\Theta(\theta) & \sim & \frac{c_c \rho_n^0}{c_1 \rho_0^2 S_0} ,
\label{eq:approx}
\eea
and ignore the possible contribution of the second term on the right hand 
side of (\ref{T2par2}). 
In this case, use of (\ref{T2par2}) and (\ref{p1par2}) leads to the 
following estimate for the ratio of energy flux scattered as first sound 
to energy flux scattered as second sound for an incident second sound 
wave: 
\begin{equation}
{I_{2\to 2} \over I_{2\to 1}} \sim \frac{\rho_0^2}{\rho_s^0 \rho_n^0} 
\left( \frac{c_1}{c_2} \right)^5
\frac{|\wbomegap(\q_{2\to2})|^2}{|\wbomegap(\q_{2\to1})|^2} .
\label{eq:estimate2}
\end{equation}
Similarly, use of (\ref{p1par1}) and (\ref{T2par1}) leads to the following 
estimate for the ratio between the energy flux scaterred as first sound 
and the energy flux scattered as second sound for an incident first sound 
wave:
\begin{equation}
{I_{1\to 1} \over I_{1\to 2}} \sim \frac{\rho_s^0}{\rho_n^0} 
\frac{c_2}{c_1} 
\frac{|\wbomegap(\q_{1\to1})|^2}{|\wbomegap(\q_{1\to2})|^2} .
\label{eq:estimate1}
\end{equation}
It is apparent that, for both incident sound modes, most of the energy 
will be 
scattered as second sound by superfluid vorticity.
A plausible explanation of this effect is as follows: the scattering is 
due to the vortex being jiggled by the incoming sound, and the 
corresponding reemission of sound. Now, the vortex 
involves superfluid motion only, so that most of the sound reemited will 
involve the superfluid velocity being out of phase with the normal 
component velocity, as appropriate for second sound waves.

 We now work out a 
specific example to validate this result.

\subsection{Scattering by a vortex dipole} 
Consider a vortex dipole oriented along the $x_2$-axis, 
build of two counter-rotating vortices of circulation $\pm \Gamma$, 
at distances $\pm a$ from the $x_1$-axis; the induced flow is not 
stationary, and the dipole moves along the $x_1$-axis at the 
constant speed $V_d = \Gamma/(4\pi a)$ \cite{lamb}. For wavelengths 
$\lambda$ large compared to the vortex separation $a$ as well as vortex 
core radius, the 
vorticity field can be approximated as
\beq
\bomega = \Gamma 2a \delta'(x_2)\delta (x_1 - V_d t)\hat x_3, 
\label{dipole}
\eeq
where $\delta (.)$ is the Dirac distribution and $\delta' (.)$ 
its derivative.
The Fourier transform of the vorticity must be calculated for a 
wave vector
\beq
q_{i\to j} = \left( \begin{array}{c} -{\nu_0\over c_i} + 
{\nu\over c_j}\cos \theta \\ {\nu\over c_j}\sin \theta \\ 0 
\end{array}\right) ; \label{qgen}
\eeq
its only nonzero component is $\wbomega_3$ and reads
\beq
\wbomega_3(\q_{i\to j}, \Delta\nu) = 
{\Gamma L \over (2\pi)^3}\delta \left[\Delta \nu - 
\left({\nu_0\over c_i} - 
{\nu\over c_j}\right)V_d\right](-2ia){\nu \over c_j}\sin \theta 
\label{dipoleFT}
\eeq
where $L$ is either the length of the dipole of the extension of the 
sound wave in the $x_3$ direction. A factor $a/\lambda_j$ appears in 
\refpar{dipoleFT}, which comes from the appearance of a lower wavelength 
limit in the problem, {\it i.e.}, the dipole dimension $a$. The following 
calculations are thus valid for long wavelength ($\lambda \gg a$), which 
reinforces the tendency already shown by \refpar{eq:estimate1} and 
\refpar{eq:estimate2}, to be compared respectively with the explicit 
calculations \refpar{ratio21sur22} and \refpar{ratio12sur11}. 

>From the expression 
\refpar{dipoleFT} we can calculate the angular dependence of the 
scattered waves, using
\bea
(\vecinc\times \hat r)\cdot \wbomega &=& \wbomega_3 \sin \theta 
\label{angle1} \\ 
\hat r_i \vecinc_j  \widetilde{\Omega}_{ij} &=& 
-\wbomega_3 \cos \theta\label{angle2}
\eea
Since our formulae are valid in three dimensions, it is assumed that any 
detectors are placed at distances large, compared with $L$, from the 
vortex dipole.

>From \refpar{p1par2} and \refpar{T2par2} we may now calculate 
the ratio of the total scattered intensity of first and second 
sound waves when the vortex dipole is illuminated by a second 
sound wave. It reads
\beq
{I_{2\to 1} \over I_{2\to 2}} = {\rho_s^0 \rho_n^0\over  \rho_0^2} 
4\left({c_2 \over c_1}\right)^7{\int_0^{2\pi}{\cos^2\theta \sin^4\theta 
/ (1 + \alpha^2 - 2\alpha\cos \theta)^2}{\rm d}\theta 
\over \int_0^{2\pi}\left(H_c(\theta)\sin^2\theta + 
(B/2)\sin \theta \cos \theta\right)^2{\rm d}\theta}  
\label{ratio21sur22debut}
\eeq
where $\alpha \equiv c_2/c_1$ is a small quantity. The integral in 
the numerator is given in the tables of Gradshteyn and Ryzhik 
\cite{gradstein} and its value is $\pi/8$ to leading order in 
$\alpha$. The numerical value of the integral in the denominator 
is about $6\pi$; thus
\beq
{I_{2\to 1} \over I_{2\to 2}} = {\rho_s^0 \rho_n^0\over  \rho_0^2} 
{1\over 12}\left({c_2 \over c_1}\right)^7 \label{ratio21sur22}
\eeq
to leading order in $c_2/c_1$; this is indeed very small, as it was 
anticipated above on the basis of 
general considerations. The tendency for most of the energy to be 
scattered 
into second sound is increased by the dipolar nature of the target, which 
implies long wavelength incident waves.

Similarly, from \refpar{T2par1} and \refpar{p1par1} we may compute 
the ratio of the total scattered intensity of first and second 
sound waves when the vortex dipole is illuminated by a second 
sound wave. The result is
\beq
{I_{1\to 2} \over I_{1\to 1}} = {\rho_n^0 \over \rho_s^0 } 
\left({c_1 \over c_2}\right)^5
{\int_0^{2\pi}\left(\Theta(\theta)\sin^2\theta\right)^2{\rm d}\theta 
\over \int_0^{2\pi}\cos^2\theta (\cos \theta + 1)^2{\rm d}\theta}  
\label{ratio12sur11debut}
\eeq
The evaluation of the integral in the numerator is easy to perform 
with the help of \cite{gradstein}; to leading order in the small quantity 
$c_2/c_1$, the ratio reads
\beq
{I_{1\to 2} \over I_{1\to 1}} = 
{1\over 7}\;{\rho_n^0 \over \rho_s^0 }\;
\left({ c_1 \over  c_2}\right)^3\left\{3 {T_0^2 
S_0^4{\rho_s^0}^2\rho_0^2\over 
c_1^2c_2^2 C^2 {\rho_n^0}^2}\left[{\rho_0 \over 
\rho_s^0}\left({\partial 
\rho_s \over \partial \rho}\right)^0_T -1\right]^2  + 2\right\} . 
\label{ratio12sur11}
\eeq
This quantity may be quite large not too far from the $\lambda$-point (the 
cumbersome expression in the brackets is of order one). Again, the 
tendency for most of the energy to be scattered as second sound 
is increased at long wavelengths in the case of a vortex dipole.

\section{Concluding Remarks}
We have studied the scattering of first and second sound waves by quantum 
vorticity in superfluid Helium within the context of two fluid 
hydrodynamics. Within a first Born approximation explicit expressions have 
been obtained for the scattering of second sound into both first and 
second sound (Eqns. (\ref{T2par2}) and (\ref{p1par2})), as well as 
for the scattering of first sound into both first and second sound 
(Eqns. (\ref{p1par1}) and (\ref{T2par1})). 
The relevant formulae coincide with cases already studied 
(second sound scattered into second sound\cite{Pitaevskii,LundSteinberg} 
and first 
sound into first sound\cite{LundRojas}) in the appropriate limits. It 
appears that experimental verification of these effects is not outside the 
possibilities of currently available hardware\cite{LundSteinberg}. In that 
case, a new tool to study quantum vorticity might become available.

\paragraph{Acknowledgements}

This work was suppported by ECOS/Conicyt,
 Fondecyt Grant 1960892 and a C\'atedra Presidencial en Ciencias.

%\newpage 

\end{document}